\documentclass[aps,pre,twocolumn,floats,amsmath,amssymb,showpacs,floatfix]{revtex4}
\usepackage{graphicx}
\usepackage{dcolumn}
\usepackage{psfrag}
\usepackage{float}

\begin{document}
\title{Distribution of relative velocities in turbulent aerosols}
\author{K. Gustavsson and  B. Mehlig}
\affiliation{Department of Physics, Gothenburg University, 41296
Gothenburg, Sweden} 
\begin{abstract}
We compute the distribution
of relative velocities for a one-dimensional model of heavy particles suspended in
a turbulent flow, quantifying the caustic contribution to the moments of relative
velocities. The same principles determine
the corresponding  caustic contribution in $d$ spatial dimensions.
The distribution of relative velocities $\mbox{\boldmath$\Delta v$}$ at small separations $R$ 
acquires the universal form $\rho(\mbox{\boldmath$\Delta v$},R) \sim R^{d-1}|\mbox{\boldmath$\Delta v$}|^{D_2-2d}$ for large (but not too large) 
values of $|\mbox{\boldmath$\Delta v$}|$. 
Here $D_2$ is the phase-space correlation dimension.
Our conclusions are in excellent agreement with numerical
simulations of particles suspended in a randomly mixing
flow in two dimensions, and in quantitative agreement with
published data on direct numerical simulations
of particles in turbulent flows.
\end{abstract}
% 05.40.-a Fluctuation phenomena, random processes, noise, and Brownian  motion
% 05.60.Cd Classical transport
% 46.65.+g Random phenomena and media
% 05.40.Jc Brownian motion
% 47.27.Qb Turbulent diffusion
% 92.60.Mt Particles and aerosols
% 45.50.Tn Collisions
% 47.27.-i Turbulent flows

\maketitle

Collisions of particles in randomly mixing or turbulent flows (\lq turbulent aerosols')
have been studied intensively for several decades. 
The main goal, not yet achieved, is to find
a reliable parameterisation of the
collision kernel. This is an important
question since the frequency of collisions
between suspended particles determines the
stability of turbulent aerosols.
Direct numerical simulations of particles in turbulent
flows (see for example \cite{Sun97,Wan00})
show that collision velocities (and thus the collision kernel)
increase precipitously
as the \lq Stokes number' ${\rm St}$  is varied beyond a threshold.
This dimensionless
parameter, ${\rm St} = (\gamma\tau)^{-1}$, is defined in terms of
the particle damping rate $\gamma$
and the relevant correlation time $\tau$ of the flow - both explained below.
In \cite{Wil06} (see also \cite{Fal02})
this steep increase was attributed to the fact that singularities
(so-called \lq caustics') in the particle dynamics result in large relative velocities at 
small separations.  Caustics occur when  phase-space manifolds
describing the dependence of particle velocity upon particle position
fold over. In the fold region, the velocity field at a given point
in space becomes multi-valued, giving rise to finite velocity
differences between particles at the same position.
In \cite{Wil06} it was argued that the collision rate
exhibits an activated ${\rm St}$-behaviour, $\exp(-A/{\rm St})$,
reflecting the frequency at which caustics occur
\cite{Dun05,Wil05,Wil06,Pum07}.

The collision kernel is determined by the distribution of relative
velocities of the suspended particles at small spatial separations. In order
to calculate this collision kernel from the microscopic equations of motion, it is necessary to understand 
how caustics determine the distribution of collision velocities, and 
to compute how caustics affect the moments of this distribution. 
This is an important question for many reasons, not only because
the ${\rm St}$-dependence of the collision rate is of fundamental
importance in aerosol physics. First, such an approach could show
which aspects of the collision dynamics in turbulent aerosols 
are universal, and which aspects depend on the particular nature
of the turbulent velocity fluctuations. Under which
circumstances are highly simplifed models appropriate
(such as a white-noise velocity field with a single
spatial scale)? Second, it is known
that turbulent aerosol particles form fractal clusters
(see \cite{Dun05} and references cited therein).
These fractal objects fold in phase space when the 
above-mentioned singularities occur. 
This raises the question: how do fractal spatial fractal clustering
and caustic singularities together determine relative velocities in turbulent aerosols?

In this paper we compute the joint steady-state distribution of
relative velocities and particle separations, first in a one-dimensional model \cite{Wil03}
for relative velocities of inertial particles using perturbation theory.
Our results show that the distribution
of relative velocities at small separations is a power law, and
that the power is determined by the correlation dimension
of the steady-state phase-space fractal. 
We show that this power law is
a consequence of caustics, and that
caustics make a substantial contribution to relative velocities
at small separations, consistent with the picture outlined above,
and the findings of Ref.~\cite{Wil06}.
Relative velocities in higher spatial dimensions
are determined by the same principles, and we show
that the corresponding power law (Eq. (\ref{eq:asymV_2d}) below) 
is a universal property 
of turbulent aerosols. It determines, in general, how caustics
and fractal clustering give rise to fractal phase-space distributions
which in turn determine the collision kernel. 
Last but not least, our findings explain the results of a recent scaling
analysis of the particle-velocity structure functions obtained
by direct numerical simulations of turbulent flows \cite{Bec10}.

\begin{figure}[t]
\includegraphics[width=8.5cm,clip]{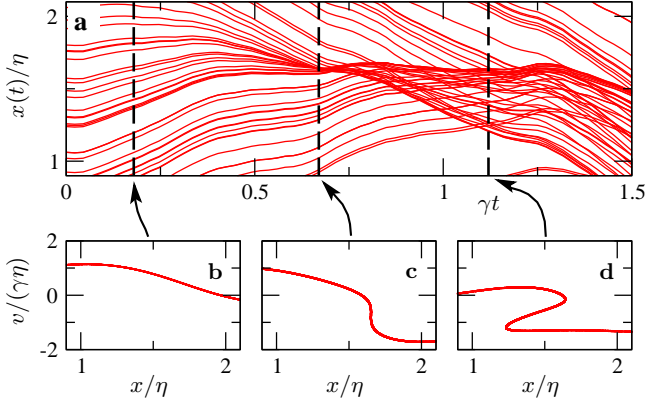}
\caption{\label{fig:1} 
{\bf a} Particle trajectories $x(t)$ according
to Eq.~(\ref{eq:deutsch}) 
with a Gaussian random field $u(x,t)$,
with $\langle u\rangle = 0$ and
$\langle u(x,t)u(0,0)\rangle=u_0^2\,
\exp\left[-x^2/(2\eta^2)-|t|/\tau\right]$
for small values of $x$, and periodic boundary conditions in $x$ with period $L$.
Parameters: $\tau=0.01$, $\eta=0.1$,
$L=1$, $u_0=1$, $\gamma=4/9$, that is ${\rm Ku} = 0.1$, ${\rm St} = 225$
(particles initially at rest).
Panels {\bf b} to {\bf d} show how the corresponding 
phase-space manifold folds.}
\end{figure}
\begin{figure}
\includegraphics[width=4.2cm,clip]{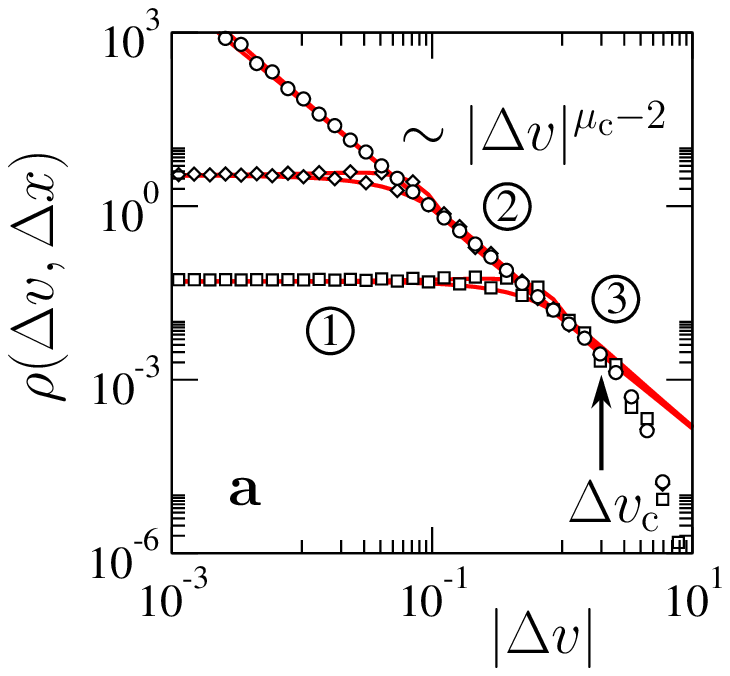}
\hspace*{6mm}\raisebox{6mm}{\includegraphics[width=3.7cm,clip]{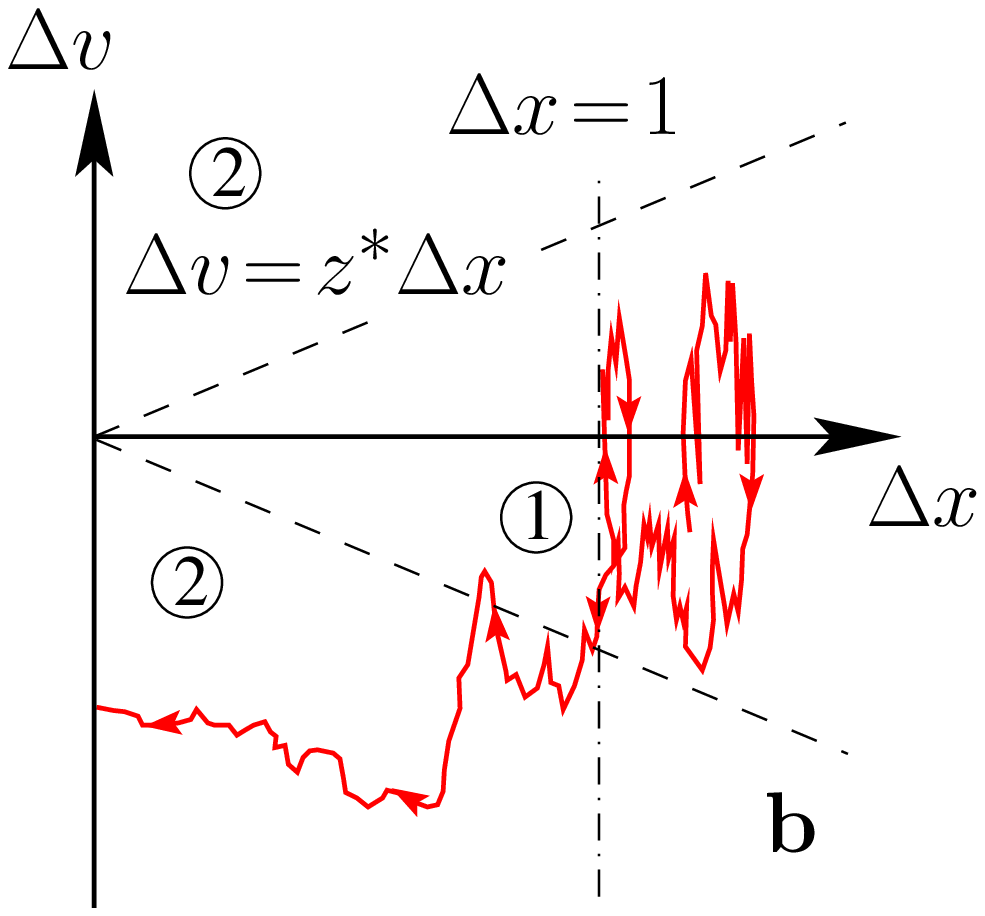}}\\[2mm]
\hspace*{2.5mm}\includegraphics[width=3.8cm,clip]{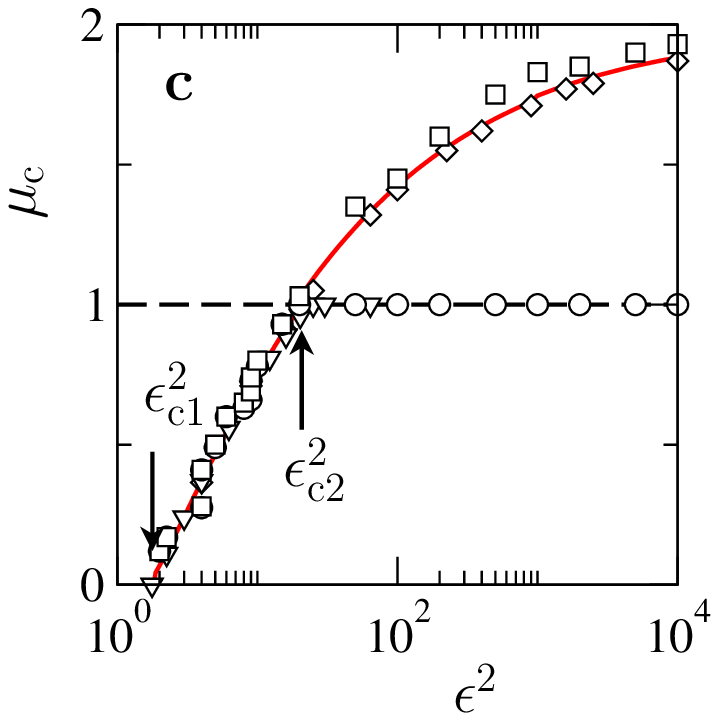}
\raisebox{-2mm}{\includegraphics[width=4.2cm,clip]{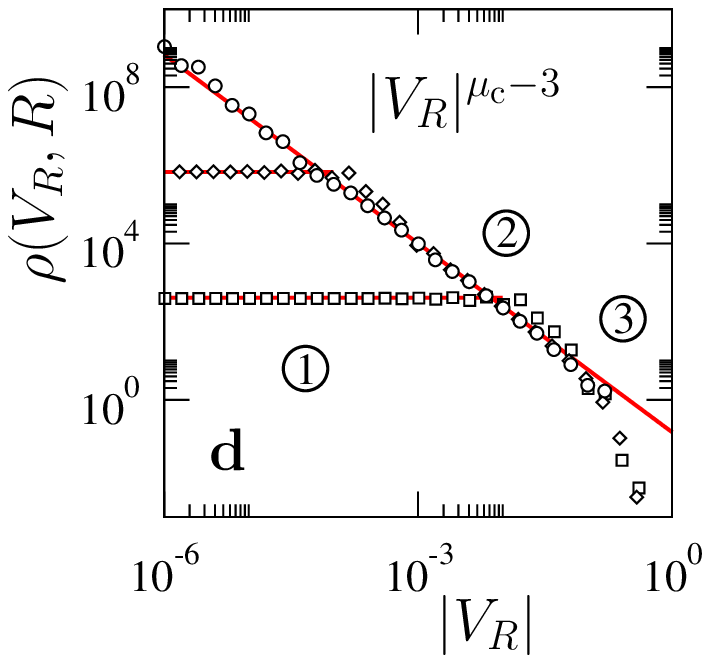}}
\caption{\label{fig:2}
{\bf a} Steady-state 
relative-velocity distribution $\rho(\Delta v,\Delta x)$ 
at $\Delta x = 6\times 10^{-4}$ ($\circ$), $5\times10^{-2}$ ($\Diamond$),
and $5 \times 10^{-1}$ ($\Box$), otherwise same parameters as in Fig.~1.  Also shown are
results from (\ref{eq:Z1}) subject to the constraint $A_\mu^+ = A_\mu^-$ (solid red lines).
{\bf b} Typical trajectory in the $\Delta x$-$\Delta v$-plane.
Arrows indicate the direction of motion.
{\bf c}
Results for $\mu_{\rm c}(\epsilon^2)$ from  (\ref{eq:Z2}), red line,
numerical results for the spatial correlation dimension $d_2$ ($\circ$),
and numerical results for the phase-space correlation dimension $D_2$ ($\Box$),
both for ${\rm Ku}=0.1$.
Also shown are results of a linearised Langevin model (${\rm Ku}=0$) for
$d_2$ ($\triangledown$) and $D_2$ ($\Diamond$). Two critical values
of $\epsilon^2$ are shown: for the path-coalescence transition \cite{Wil03}, 
$\epsilon^2_{\rm c1} \approx 1.77$, and the second critical value
$\epsilon^2_{\rm c2}\approx 20.7$  where $d_2$ equals unity.
{\bf d} Numerical results (symbols) for
the distribution $\rho(V_R,R)$ of relative velocities from computer simulations 
of a linearised Langevin model
for inertial particles suspended in
a two-dimensional incompressible Gaussian random flow, $\epsilon^2 = 0.04$,
at $R = 5\times 10^{-6}$ ($\circ$), 
$5 \times 10^{-4}$ ($\Diamond$), and $5\times 10^{-2}$ ($\Box$). 
Also shown is the theoretical  result Eq.~(8) (solid red line), the value of $\mu_{\rm c}=D_2$
was determined by numerically computing the phase-space correlation dimension $D_2$.
}
\end{figure}

We first analyse the one-dimensional model \cite{Deu85,Wil03}
\begin{equation}
\label{eq:deutsch}
\ddot x = \gamma (u(x,t)-v)\,.
\end{equation}
Dots denote time derivatives, $v = \dot x$ is the particle velocity,
$\gamma$ the rate at which the inertial
motion is damped relative to the fluid, and $u(x,t)$ is
a Gaussian random field describing the velocity of the fluid.
Fig.~\ref{fig:1} shows particle trajectories
according to (\ref{eq:deutsch}) and illustrates
how a smooth manifold of initial conditions develops
fold singularities (caustics).
We assume that $u(x,t)$  is characterised
by spatial and temporal correlation scales $\eta$ and $\tau$,
and by its typical size, $u_0$
(the statistical properties of $u(x,t)$ 
are described in the caption to Fig.~\ref{fig:1}). 
The dynamics is then determined by two dimensionless parameters,
by the Stokes number ${\rm St}$ and by the
 \lq Kubo number' ${\rm Ku} = u_0\tau/\eta$ which
characterises the  fluctuations of $u(x,t)$  \cite{Dun05,vKa81}.
In turbulent flows the Kubo number is of order unity, but
in the following we analyse the dynamics in the limit
of small values of ${\rm Ku}$ where  the
suspended particles experience $u(x,t)$ as a white-noise
signal. 
In the limit ${\rm Ku}\rightarrow 0$, the joint density $\rho(\Delta v,\Delta x)$
is determined by a Kramers equation \cite{vKa81,Gus08} 
with a $\Delta x$-dependent diffusion constant.
In dimensionless units ($t' = \gamma t, x' =  x/\eta,  
v' = v/(\eta\gamma), u' = u/(\eta\gamma)$ and dropping the primes to
simplify the notation) we have:    
\begin{eqnarray}
\nonumber    
\partial_t\rho 
&=&-\partial_{\Delta x}
(\Delta v\,\rho)+\partial_{\Delta v}(\Delta v\,\rho)+{\mathcal D}(\Delta x)\partial^2_{\Delta v}\rho\,,\\
{\mathcal D}(\Delta x) &=& \frac{1}{2} 
\int_{-\infty}^\infty\!\!\!{\rm d}t \langle\Delta u(\Delta x,t)
\Delta u(\Delta x,0)\rangle\,.
\label{eq:FP}
\end{eqnarray}
Here $\Delta u(\Delta x,t) = u(x+\Delta x,t)-u(x,t)$. For the model
described in Fig.~\ref{fig:1} we have ${\mathcal D}(\Delta x) \sim \epsilon^2 \Delta x^2$
for $|\Delta x|\ll 1$ and ${\mathcal D}(\Delta x) \sim  2 \epsilon^2\equiv 
{\mathcal D}_0$ for $|\Delta x|\gg 1$.
Here $\epsilon^2 = {\rm Ku}^2 {\rm St}$.

In the limit of $\Delta x \rightarrow 0$, the steady-state solution of (\ref{eq:FP}) is found by separation of 
the variables $\Delta x$ and $z = \Delta v/\Delta x$. Inserting the 
ansatz $g_\mu(\Delta x) Z_\mu(z)$ with separation
constant $\mu$ into (\ref{eq:FP}) results in
$g_\mu(\Delta x)=|\Delta x|^{\mu-1}$, while $Z_\mu(z)$ solves
$0=-\mu zZ_\mu(z)+\partial_z(z+z^2+\epsilon^2\partial_z)Z_\mu(z)$.

The steady-state solution of (\ref{eq:FP}) is given by
a weighted sum of $g_\mu Z_\mu$ over the allowed values of $\mu$.
We know that the distribution of spatial separations exhibits
a power law as $\Delta x \rightarrow 0$, corresponding to spatial clustering.
We expect that the dominant contribution to this distribution derives from the smallest positive
allowed value of $\mu$. 
For certain values of $\mu$, 
the functions $Z_\mu(z)$  are known in closed form 
(for $\mu=0$ \cite{Wil03} and for $\mu= -1$ \cite{Der06}).
For other values of $\mu$, we determine an analytical 
solution by expanding $Z_\mu$ around $\mu=0$. We find:
\begin{eqnarray}
\label{eq:Z1}
Z_\mu(z) &=& \sum_{k=0}^\infty
\Big(\frac{\mu}{\epsilon^2}\Big)^k \int_{-\infty}^0
\!\!\!{\rm d}t_1\cdots {\rm d}t_{2k+1}
\Big(\prod_{i=1}^k\sum_{j=0}^{2i}t_j\Big)\nonumber\\
&\times&\exp\Big(-\!\sum_{i=0}^{2k+1}(-1)^iV(\sum_{j=0}^it_j)\Big)
\label{eq:Z2}
\end{eqnarray}
with $V(y) = \epsilon^{-2}(y^2/2+y^3/3)$ and $t_0=z$. A related expansion
was used in \cite{Sch02} to calculate moments of the  finite-time Lyapunov exponent for
particles accelerated in a random time-dependent potential.
To determine the allowed values of $\mu$ we consider
the large-$z$ asymptotes of $Z_\mu(z)$. The known solution $Z_0(z)$
\cite{Wil03} exhibits power-law tails $\sim |z|^{-2}$ reflecting the fact
that caustic singularities 
are reached in finite time. 
Correspondingly, when $z$ is large we can neglect the term $\partial_z(z Z_\mu)$
in the equation for $Z_{\mu}$. 
The resulting equation is solved by combinations of Kummer functions with the 
asymptotic behaviour $Z_\mu\sim |z|^{\mu-2}$ for large values of $|z|$. 
A power-law ansatz in the equation for $Z_\mu$ gives
\begin{equation}
\label{eq:asy}
Z_\mu(z)\sim A^\pm_\mu\, (\pm z)^{\mu-2}( 1+(\mu-1)z^{-1}+\ldots)
\end{equation}
for $z \rightarrow  \pm \infty$.  Eq.~(\ref{eq:asy})  suggests that 
$\rho(\Delta v, \Delta x)$ approaches power-law form 
as $\Delta x\rightarrow 0^+$
\begin{equation}
\label{eq:xz}
\rho(\Delta v, \Delta x) 
=|\Delta x|^{\mu-2} Z_{\mu}({\Delta v}/{\Delta x})
\sim A^\pm_\mu\,(\pm \Delta v)^{\mu-2}
\end{equation}
for positive and negative values of $\Delta v$, respectively. 
Now we use that interchanging the particles in 
a particle pair ($\Delta x \rightarrow -\Delta x$ and
$\Delta v \rightarrow -\Delta v$) cannot alter
the distribution. In particular this must be true
at $\Delta x=0$ which requires $A^+_\mu = A^-_\mu$.
The allowed values of $\mu$ can be computed
by evaluating (\ref{eq:Z2}) and imposing that
$A_\mu^+ = A_\mu^-$. To simplify the problem we only consider the
smallest positive allowed value of $\mu$ (referred to as $\mu_{\rm c}$,
and shown in Fig.~\ref{fig:2}{\bf c}). 
This yields an accurate description (except
in the very far tails) of the distribution 
$\rho(\Delta v, \Delta x)$ for small values of $\Delta x$ as Fig.~\ref{fig:2}{\bf a} shows.

Three regions can be distinguished in $\rho(\Delta v, \Delta x)$.
First, the body of the distribution (labeled $1$ in 
Fig.~\ref{fig:2}{\bf a}) corresponds to small values
of $z$ ($|z|\ll 1$ that is $|\Delta v| \ll |\Delta x|$).
Typically, the separation remains constant in this region, $\Delta x \approx \Delta x_0$,
and $\Delta v$ obeys ${\rm d}\Delta v/{\rm d}t \approx -\Delta v + \Delta u$. In the limit ${\rm Ku}\rightarrow 0$,
${\rm St}\rightarrow \infty$ so that $\epsilon^2 = {\rm Ku}^2 {\rm St}$ remains
constant,
this is an Ornstein-Uhlenbeck process. It gives rise
to a broad Gaussian distribution of relative velocities,
approximately independent of $\Delta v$.
This smooth contribution is not universal.

Second, the regime $|z|\gg 1$ (that is $|\Delta v|\gg |\Delta x|$, labeled $2$ in
Fig.~\ref{fig:2}{\bf a}) corresponds to particles 
approaching each other on different caustic branches. 
This singular contribution, due to caustics,
yields large values of $|\Delta v|$ as $|\Delta x| \rightarrow 0$.
In this case, 
the deterministic part of Eq.~(\ref{eq:FP}) dominates, 
giving rise to 
linear trajectories $\Delta v = \Delta v_0+\Delta x_0-\Delta x$.
When $|\Delta v|\gg |\Delta x|$, we have $\Delta v \approx \Delta v_0$.
In this limit, the distribution $\rho(\Delta v, \Delta x)$ becomes
approximately independent of $\Delta x$. As shown above,
it exhibits the power-law form (\ref{eq:xz}). The
power-law is clearly visible in Fig.~\ref{fig:2}{\bf a}. 
This singular caustic contribution to $\rho(\Delta v, \Delta x)$
gives rise to large moments of $\Delta v$, 
and thus to a large collision rate. We emphasise that the dynamics
in this regime is universal, that is independent of the particular
statistics of the velocity field.
Fig.~\ref{fig:2}{\bf b} shows a typical phase-space trajectory passing through the regimes $1$ and $2$ and contributing
to the smooth and singular parts of the distribution $\rho(\Delta v,\Delta x)$.

Third, as Fig.~\ref{fig:2}{\bf a} shows, the power law is cut off
at very large velocities, ensuring 
that the moments of $\Delta v$ do not diverge.  This region (labeled $3$ in
Fig.~\ref{fig:2}{\bf a}) is not described by keeping the smallest
positive value of $\mu_{\rm c}$ only.
The cut off, $\Delta v_{\rm c}$, is simply due to the fact that the
relative velocities $\Delta v_0$ at 
large separations are of order ${\mathcal D}_0^{1/2}$. 
This suggests that $\Delta v_{\rm c} \sim  {\mathcal D}_0^{1/2}$.
The form of $\rho(\Delta v, \Delta x)$ at $|\Delta v| \gg \Delta v_{\rm c}$ 
depends upon the details of the model. 
In turbulent flows at large Stokes and Reynolds numbers, $\Delta v_{\rm c}$ is 
determined by the \lq variable-range projection principle' 
suggested in \cite{Gus08}.
The cut off $\Delta v_{\rm c}$  corresponds to a cut off $\Delta v_{\rm c}/\Delta x$ in the distribution of $z$. 
Thus 
$\rho(\Delta x,z)$ does not strictly factorise, except in the limit $\Delta x\rightarrow 0$.
The probability distribution of $z$ for large values of $|z|$ 
is obtained by integrating the joint distribution over $\Delta x$ to $\pm\Delta v_{\rm c}/|z|$. As a consequence of the power law  (\ref{eq:asy}), 
the tails of $Z_0(z)$ found in \cite{Wil03} are recovered, namely
$Z_0(z)  \sim |z|^{-2}$.
This shows once 
more that the distribution of the finite differences $\Delta v$ and $\Delta x$
is governed by the same principles as the fluctuations of $\partial v/\partial x$.

What do these results imply for the moments
$m_p(\Delta x) = \int\!{\rm d}\Delta v |\Delta v|^p \rho(\Delta v, \Delta x)$
at small $\Delta x$? We find:
\begin{eqnarray}
\nonumber
m_p(\Delta x) &=& |\Delta x|^{p+\mu_{\rm c}-1} \int\!\!{\rm d}z |z|^p Z_{\mu_{\rm c}}(z)\\
&&\hspace*{-1.5cm}\sim |\Delta x|^{p+\mu_{\rm c}-1}(b_p + c_p |\Delta x|^{-p-\mu_{\rm c}+1})\,.
\label{eq:z}
\end{eqnarray}
Here the coefficient $b_p$ results from the contribution of the body of $Z_{\mu_{\rm c}}(z)$
to the $z$-integral in (\ref{eq:z}). The second term is the caustic contribution,
it  results from integrating the power-law tails of $Z_{\mu_{\rm c}}(z)$ up to the cut-off $\Delta v_{\rm c}/\Delta x$,
keeping only the leading-order behaviour in (\ref{eq:asy}). 
This form of $m_p(\Delta x)$ was 
deduced from results of numerical simulations for the distribution of relative velocities and separations in 
a one-dimensional Kraichnan model \cite{Cencini}, and it was suggested
that the exponent $\mu_{\rm c}$ equals the spatial correlation dimension.
This is correct for $\mu_{\rm c} < 1$, as the following argument shows.
The spatial correlation dimension $d_2$
describes the power-law scaling for $p=0$, namely $m_0(\Delta x)\sim |\Delta x|^{d_2-1}$ 
as $\Delta x \rightarrow 0$. From Eq.~(\ref{eq:z}) we deduce
$m_0(\Delta x) \sim |\Delta x|^{{\rm min}(\mu_{\rm c},1)-1}$,
implying that $\mu_{\rm c}=d_2$ for $0 \leq \mu_{\rm c} < 1$.
But for $\mu_{\rm c} > 1$ this is not the case. A change of variables $\Delta x \rightarrow 
\Delta w = \sqrt{\Delta x^2+\Delta v^2}$ in (\ref{eq:xz}) 
allows us to compute the distribution of \lq phase-space separations' $\Delta w$.
We find that it exhibits a power-law of the form $\Delta w^{\mu_{\rm c}-1}$ as $\Delta w\rightarrow 0$. 
This proves that $\mu_{\rm c}$ is in fact equal to the phase-space
correlation dimension $D_2$ (which ranges between $0$ and $2$).
It also demonstrates that $D_2 = d_2$ for $0 \leq D_2  \leq 1$ (consistent with
results of numerical simulations of particles in two-dimensional random flows at finite Kubo numbers \cite{Bec}).
We computed spatial and phase-space correlation dimensions
from numerical simulations of the equation of motion (\ref{eq:deutsch}). The results
are shown in Fig.~\ref{fig:2}{\bf c} and are in good
agreement with the analytical theory.
Caustic contributions dominate for $p > 1-\mu_{\rm c}$, the smooth contribution
is thus irrelevant for $p=1$. 

For two- and three-dimensional
incompressible flows, the distribution of relative velocites $\mbox{\boldmath$\Delta v$}$ at spatial separation $R$
has a form similar to (\ref{eq:xz}): as $R\rightarrow 0$, the $d$-dimensional 
analogue of 
equation (\ref{eq:FP}) is solved by the ansatz 
$g_\mu(R)Z_\mu(\mbox{\boldmath$z$})$ where $\mbox{\boldmath$z$}=\mbox{\boldmath$\Delta v$}/R$. We find that $Z_\mu$ exhibits tails of the form
$Z_\mu(\mbox{\boldmath$z$})\sim |\mbox{\boldmath$z$}|^{\mu-2d}$. Comparison with the expected form 
of the distribution of (phase-)space separations demonstrates that $\mu_{\rm c}$ equals the
phase-space correlation dimension. We find:
\begin{equation}
\rho(\Delta\mbox{\boldmath$v$},R)\sim R^{d-1}|\Delta\mbox{\boldmath$v$}|^{D_2-2d}\,.
\label{eq:asymV_2d}
\end{equation}
Now consider the 
distribution of relative 
radial velocities $V_R\equiv \Delta\mbox{\boldmath$v$}\cdot\hat R$
($\hat R$ is the radial unit vector). For $|V_R| \gg R$ this distribution is 
found by integration of (\ref{eq:asymV_2d}). Assuming $D_2 < d+1$,
we find for small values of $R$ (and $|V_R| \gg R$)
\begin{equation}
\rho(V_R,R)\sim R^{d-1}|V_R|^{D_2-d-1}\,.
\label{eq:asymVr_2d}
\end{equation} 
Eq.~(\ref{eq:asymVr_2d}) yields the following
expression for the moments of $m_p(R)$ of $V_R$:
\begin{equation}
\label{eq:mp}
m_p(R) =b_p({\rm St}) R^{p+D_2-1} + c_p({\rm St}) R^{d-1}\,,
\end{equation}
of the same form as (\ref{eq:z}). But note that here 
the singular caustic term acquires a geometrical factor $R^{d-1}$.
Setting $p=0$ in (\ref{eq:mp}) shows 
that $d_2 = \min\{D_2,d\}$ in analogy with the one-dimensional case.
For $p=1$, Eq.~(\ref{eq:mp}) is equivalent
to the ansatz proposed in \cite{Wil06}, apart from the  difference that in
Eq.~(7) of \cite{Wil06}, 
the asymptotic behaviour $D_2({\rm St})\rightarrow d$
as ${\rm St}\rightarrow 0$ was used for the first term in (\ref{eq:mp}).

Fig.~\ref{fig:2}{\bf d} shows numerical results for
a linearised Langevin model for inertial particles
in a two-dimensional incompressible Gaussian random
flow.  Shown is the distribution $\rho(V_R,R)$.
As in the one-dimensional case, the power-laws in (\ref{eq:asymV_2d}) and
(\ref{eq:asymVr_2d}) represent the universal contribution of caustics
to the distribution of relative velocities at small separations.
This demonstrates how caustics and fractal clustering determine
the distribution of relative velocities at small separations.
Our numerical results imply that
the prefactor $c_p({\rm St})$ of the singular caustic term
in (\ref{eq:mp}) is of activated form
$c_p({\rm St}) \sim {\rm e}^{-A_p/\epsilon^2}$ for small $\epsilon$ (not shown).

In the remainder we show how our theory explains
the results of direct numerical simulations of particles
suspended in turbulent flows  \cite{Bec10}.
In \cite{Bec10}, particle-velocity structure functions $S_p(R) = m_p(R)/m_0(R)$
were computed numerically and analysed in terms
of the scaling ansatz $S_p(R) \sim R^{\xi_p}$.
Fig.~3 in \cite{Bec10} shows the scaling
exponent $\xi_1$ as a function of ${\rm St}$.
Our theory, Eq.~(\ref{eq:mp}), predicts
that $\xi_1 \approx 1$ for small values of ${\rm St}$, and $\xi_1 = d-d_2$ for large values of ${\rm St}$.
These  predictions are in quantitative agreement with the results
presented in Fig.~3 in \cite{Bec10}.
Our theory also explains the distribution of $z_R = v_R/R$
shown in Fig.~5 in \cite{Bec10a}. Eq.~(\ref{eq:asymVr_2d})
yields $P(z_R|R) \sim |z_R|^{D_2-d-1}$ which 
explains the power law in the inset of Fig.~5 in \cite{Bec10}.
The far tails in this figure are expected to
be determined by the variable-range projection
principle \cite{Gus08} yielding $P(z_R|R) \sim \exp(-C\,|z_R\, R|^{4/3})$.
As a next step it is necessary to determine whether the above expression
describes the functional form of the far tails of this
distribution in \cite{Bec10a}.

%{\em Acknowledgements}. 
Financial support by Vetenskapsr\aa{}det, by the G\"oran Gustafsson Stiftelse, 
and by the platform for \lq Nanoparticles in interactive environments' 
is gratefully acknowledged.

\end{document}